\documentclass{article}

\usepackage{spconf,amsmath,graphicx,hyperref}
\usepackage{booktabs}
\usepackage{multirow}
\usepackage{array}
\usepackage{makecell}
\usepackage{amssymb}
\usepackage{adjustbox}
\usepackage{caption}

\makeatletter
\def\name#1{\gdef\@name{{\em #1}\par}}  
\def\@maketitle{\newpage
  \null
  \vskip 2em \begin{center}
  {\large \bf \@title \par} \vskip 1.5em 
  {\large \lineskip .5em
    \@name 
    \@address
  \par} \end{center}
  \par
  \vskip 1.5em}
\makeatother

\title{
    MeanFlowSE: One-Step Generative Speech Enhancement via MeanFlow
}

\name{
  Yike Zhu$^1$$~^\dagger$, Boyi Kang$^{2,1}$$~^\dagger$, Ziqian Wang$^1$, Xingchen Li$^{1,3}$, Zihan Zhang$^3$, Wenjie Li$^3$, \newline
  Longshuai Xiao$^3$, Wei Xue$^2$, Lei Xie$^1$$~^*$
}

\address{
    $^{1}$Audio, Speech and Language Processing Group (ASLP@NPU), School of Software, 
        \\ Northwestern Polytechnical University, Xi’an, China \\
    $^{2}$The Hong Kong University of Science and Technology, Hong Kong, China \\
    $^{3}$Huawei Technologies, China \\
    \texttt{ykzhu@mail.nwpu.edu.cn, lxie@nwpu.edu.cn, bkangaa@connect.ust.hk}
}

\begin{document}
\maketitle

\let\oldthefootnote\thefootnote               
\renewcommand{\thefootnote}{$\dagger$}        
\footnotetext{Equal contribution.}
\renewcommand{\thefootnote}{$*$}              
\footnotetext{Corresponding author.}
\renewcommand{\thefootnote}{\oldthefootnote}  

\begin{abstract}
Speech enhancement (SE) recovers clean speech from noisy signals and is vital for applications such as telecommunications and automatic speech recognition (ASR). While generative approaches achieve strong perceptual quality, they often rely on multi-step sampling (diffusion/flow-matching) or large language models, limiting real-time deployment. To mitigate these constraints, we present \textbf{MeanFlowSE}, a one-step generative SE framework. It adopts MeanFlow to predict an average-velocity field for one-step latent refinement and conditions the model on self-supervised learning (SSL) representations rather than VAE latents. This design accelerates inference and provides robust acoustic–semantic guidance during training. In the Interspeech 2020 DNS Challenge blind test set and simulated test set, MeanFlowSE attains state-of-the-art (SOTA) level perceptual quality and competitive intelligibility while significantly lowering both real-time factor (RTF) and model size compared with recent generative competitors, making it suitable for practical use. The code will be released upon publication at~\url{https://github.com/Hello3orld/MeanFlowSE}.
\end{abstract}


\section{Introduction}
\label{sec:introduction}

\begin{figure*}[ht]
    \centering
    \includegraphics[width=0.7\linewidth]{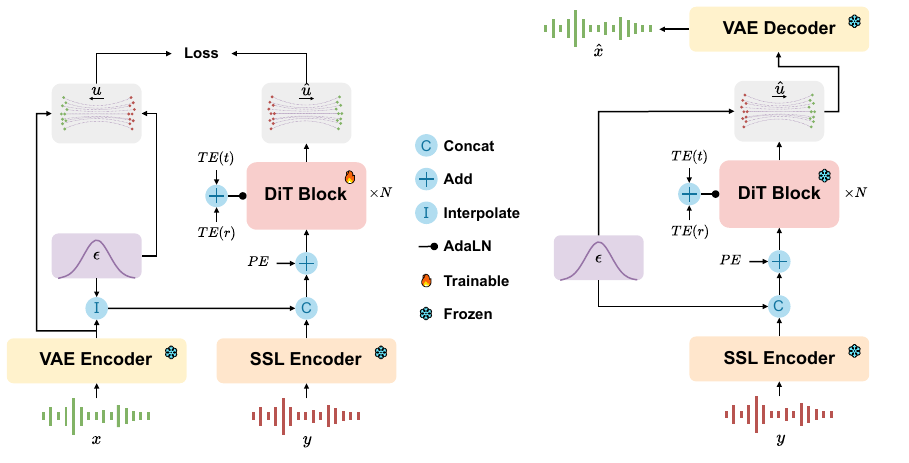}
    \caption{Overview of the proposed MeanFlowSE architecture. The left side depicts the training pipeline, while the right side illustrates the one-step inference procedure.}
    \label{fig:architecture}
\end{figure*}

Speech enhancement (SE) aims to remove interference from noisy signals and is essential for applications such as telecommunications, hearing aids, and automatic speech recognition (ASR). In recent years, neural network (NN) based methods have greatly improved the performance of speech enhancement. NN-based SE methods fall broadly into two paradigms: discriminative and generative. Discriminative methods aim to directly estimate clean speech or a corresponding mask from noisy inputs. Although effective in matched environments, such approaches often exhibit limited generalizability to unseen acoustic conditions and are prone to introduce artifacts or distortions, especially in low signal-to-noise ratio (SNR) scenarios~\cite{wang2023tfgridnetmakingtimefrequencydomain,lu2023mp,zhao2022frcrn}.

Generative methods, in contrast, aim to model the distribution of clean speech and reconstruct it through probabilistic frameworks, such as diffusion models~\cite{CDiffSE,StorM,richter2023sgmse}, language models (LM)~\cite{wang2024selm, yao2025gensegenerativespeechenhancement,mu2025fromcontinous}~and masked transformer~\cite{li2024masksrmaskedlanguagemodel,zhang2025anyenhance}~based approaches. Recently, generative modeling has achieved remarkable success in SE, producing high-quality speech reconstruction and demonstrating strong robustness under low SNR conditions. Nevertheless, the substantial computational resources required by such models, particularly the multiple sampling steps in diffusion/flow-matching-based methods~\cite{zheng2025onestepdiffusion,lipman2023flowmatchinggenerativemodeling}~greatly hinder their deployment on low resource devices. Moreover, current diffusion/flow-matching-based SE models typically condition only on noisy mel-spectrograms or latents of the noisy waveform from variational autoencoders (VAE)~\cite{wang2025flowse}, which provide irregular and noisy representations, thereby leading to suboptimal naturalness and intelligibility in the generated outputs.

To mitigate the inefficiency of multi-step sampling, recent studies have explored one-step generative frameworks~\cite{zheng2025onestepdiffusion,yin2024onestepdiffusiondistributionmatching,xu2025robustonestepspeechenhancement}. Among them, MeanFlow~\cite{geng2025mean}~introduces a principled formulation that leverages the average velocity, defined as the displacement over a time interval, instead of the instantaneous velocity in standard flow-matching. This reformulation links average and instantaneous velocities, offering a stable and efficient training target for one-step generation. Though still trailing multi-step models, MeanFlow demonstrate superior performance than previous one-step generation frameworks. Meanwhile, unlike fully generative tasks, SE benefits from the presence of a reference signal, making the problem more tractable and well-suited for one-step generative modeling.

Motivated by these insights, we propose \textbf{MeanFlowSE}, an efficient one-step generative SE framework. MeanFlowSE integrates one-step generative modeling of MeanFlow with conditioning from SSL representations, which provide fine-grained acoustic cues and rich semantic information to guide generation. Experiments show that MeanFlowSE not only achieves SOTA-level performance on public and simulated test sets but also substantially reduces computational demands, marking a significant step toward practical deployment of generative SE.

\section{Proposed Method}
\label{sec:proposed_method}

MeanFlowSE aims to recover a clean signal $x \in \mathbb{R}^T$ from a noisy signal $y \in \mathbb{R}^T$ by modeling their transformation in the latent space. As illustrated in Fig.~\ref{fig:architecture}, the framework contains three major components: (1) an SSL model to extract latent acoustic and semantic representations, (2) a latent Diffusion Transformer (DiT)~\cite{peebles2023scalable}-based MeanFlow module that predicts the average velocity field, and (3) a VAE decoder for waveform reconstruction. This modular design allows us to integrate powerful pre-trained models while keeping inference simple and efficient.

\subsection{MeanFlow for Speech Enhancement}
\label{ssec:meanflow_for_speech_enhancement}

Conventional flow-matching learns an instantaneous velocity field $v(z_t,t)$ that governs the evolution of latent states through the ordinary differential equation (ODE):
\begin{equation}
    \label{eq:flow_matching_ode}
    \frac{dz}{dt} = v(z_t,t), \quad z_0 = y, \, z_1 = x,
\end{equation}
where $z_t$ denotes the latent representation at time step $t$. During inference, solving this ODE requires multiple integration steps, which hinders real-time deployment.

To address this, MeanFlow replaces the instantaneous velocity field with an average velocity field:
\begin{equation}
    \label{eq:average_velocity_field}
    u(z_t,r,t) = \frac{1}{t-r} \int_r^t v(z_\tau,\tau)\, d\tau,
\end{equation}
which summarizes the overall trend of transformation between two time steps $(r,t)$. This formulation enables direct recovery of the clean latents in one-step:
\begin{equation}
    \label{eq:meanflow_transformation}
    z_0 = z_1 - u(z_1,0,1),
\end{equation}
where in this paper $z_1$ denotes the noisy latents and $z_0$ denotes the predicted clean latents. By learning $u$ instead of $v$, the model bypasses iterative ODE solvers~\cite{chen2019odesolver}~and achieves efficient one-step generative enhancement.

\subsection{Network Architecture}
\label{ssec:network_architecture}

As illustrated in Fig.~\ref{fig:architecture}, MeanFlowSE consists of three key modules: an SSL encoder, a VAE encoder–decoder, and a DiT-based MeanFlow backbone. Together, they provide semantic conditioning, establish a structured latent domain, and enable efficient one-step generative enhancement.

\subsubsection{SSL Encoder}
\label{sssec:ssl_encoder}

MeanFlowSE leverages a pre-trained SSL model to extract high-level acoustic and semantic latents $z_y$ from the noisy input $y$. These latents are used as conditioning signals for the generative backbone. Compared to noisy latents from a VAE encoder, SSL latents capture long-range dependencies and phonetic content~\cite{vectokspeech}, which are crucial for preserving intelligibility in low-SNR conditions. During inference, the SSL encoder acts as the sole feature extractor from noisy speech.

\subsubsection{VAE Encoder–Decoder}
\label{sssec:vae_encoder_decoder}

The VAE encoder–decoder establishes the latent space in which enhancement is performed. During training, the encoder maps the clean waveform $x$ into its latent representation $z_x = \text{Enc}(x)$, which serves as the supervision target. At inference, the decoder transforms the predicted clean latent $z_0$ from the generative backbone into the waveform domain, producing the enhanced waveform $\hat{x} = \text{Dec}(z_0)$. 


\subsubsection{DiT-based MeanFlow Backbone}
The DiT backbone serves as the generative core of MeanFlowSE. It takes interpolated latents $z_t$ (constructed by combining $z_x$ with Gaussian noise $\epsilon$) concatenated with noisy latents $z_y$, further enriched with positional embedding $PE$. Meanwhile, time step embeddings $\text{TE}(r)$ and $\text{TE}(t)$ are summed and injected into the DiT blocks through adaptive layer normalization (AdaLN), explicitly controlling the MeanFlow dynamics. The network is trained to predict the target average velocity field $u$, derived analytically from $z_x$ and $\epsilon$. At inference, with $(r=0,t=1)$, the model directly refines Gaussian noise $\epsilon$ into enhanced clean latents $z_0$ in one-step using the predicted average velocity field $\hat{u}$:
\begin{equation}
    \label{eq:obtain_enhanced_latent}
    z_0 = \epsilon - \hat{u}.
\end{equation}

\subsection{Training Objective}
\label{ssec:training_objective}

To optimize the MeanFlow module, we minimize the gap between predicted and target average velocities. Instead of a plain $\ell_2$ loss, we adopt an adaptive $\ell_2$ loss that reweights training samples according to their error magnitude:
\begin{equation}
    \label{eq:loss_function}
    \mathcal{L} = \mathbb{E}\!\left[ w \cdot \| \hat{u}(z_t,r,t)-u \|_2^2 \right],
\end{equation}
where $w = ( \delta^2 + c )^{-(1-\gamma)}$ is a weight depending on the sample error $\delta^2 = \| \hat{u} - u \|_2^2$, with hyperparameters $\gamma$ and $c$ controlling the adaptivity. This loss down-weights outliers (large errors) and emphasizes reliable samples, leading to more stable training and better generalization than uniform $\ell_2$.

\section{Experiments}
\label{sec:experiments}

\captionsetup[table]{aboveskip=0pt, belowskip=10pt}
\begin{table*}[ht]
    \caption{DNSMOS scores on the Interspeech 2020 DNS Challenge blind test set. ``D" denotes Discriminative, ``G" denotes Generative, ``FM($\cdot$)" denotes flow-matching with $\cdot$ inference steps.}
    \label{tab:dnsmos_comparison}
    \centering
    \begin{adjustbox}{max width=\textwidth}
    \begin{tabular}{l c ccc ccc ccc}
        \toprule[1pt]
        \multirow{2}{*}{~~~~~~~~~~~~~System}
        & \multirow{2}{*}{Type}  
        & \multicolumn{3}{c}{With Reverb} 
        & \multicolumn{3}{c}{Without Reverb} 
        & \multicolumn{3}{c}{Real Recording} \\
        \cmidrule(lr){3-5} \cmidrule(lr){6-8} \cmidrule(lr){9-11}
        & & SIG~$\uparrow$ & BAK~$\uparrow$ & OVRL~$\uparrow$ 
          & SIG~$\uparrow$ & BAK~$\uparrow$ & OVRL~$\uparrow$ 
          & SIG~$\uparrow$ & BAK~$\uparrow$ & OVRL~$\uparrow$ \\
        \midrule
        Noisy                                 & —                     & 1.760          & 1.497          & 1.392           & 3.392          & 2.618          & 2.483           & 3.053          & 2.509          & 2.255           \\
        \midrule
        FRCRN~\cite{zhao2022frcrn}            & D                     & 2.933          & 2.924          & 2.279           & 3.574          & 4.154          & 3.331           & 3.371          & 3.978          & 3.037           \\
        MP-SENet~\cite{lu2023mp}              & D                     & 2.914          & 3.444          & 2.437           & 3.595          & 4.177          & 3.374           & 3.454          & 4.046          & 3.163           \\
        \midrule
        SELM~\cite{wang2024selm}              & G                     & 3.160          & 3.577          & 2.695           & 3.508          & 4.096          & 3.258           & 3.591          & 3.435          & 3.124           \\
        AnyEnhance~\cite{zhang2025anyenhance} & G                     & 3.500          & 4.040          & 3.204           & 3.640          & 4.179          & 3.418           & 3.488          & 3.977          & 3.161           \\
        FlowSE~\cite{wang2025flowse}          & G                     & 3.614          & 4.110          & 3.340           & 3.690          & \textbf{4.200} & 3.451           & \textbf{3.643} & 4.100          & 3.271           \\
        LLaSE-G1~\cite{kang2025llase}         & G                     & 3.594          & 4.096          & 3.334           & 3.661          & 4.173          & 3.415           & 3.472          & 3.996          & 3.177           \\
        \midrule
        MeanFlowSE                            & G                     & 3.615          & \textbf{4.177} & 3.368           & 3.668          & 4.183          & 3.438           & 3.564          & \textbf{4.139} & 3.298           \\
        MeanFlowSE$_\text{VAE input}$         & G                     & 3.101          & 4.003          & 2.727           & 3.492          & 4.131          & 3.226           & 3.320          & 4.029          & 3.000           \\
        MeanFlowSE$_\text{FM(40)}$            & G                     & 3.662          & 4.095          & 3.346           & 3.699          & 4.153          & 3.439           & 3.619          & 4.031          & 3.273           \\
        MeanFlowSE$_\text{FM(100)}$           & G                     & \textbf{3.681} & 4.167          & \textbf{3.420}  & \textbf{3.704} & 4.191          & \textbf{3.475}  & 3.628          & 4.117          & \textbf{3.337}  \\
        \bottomrule[1pt]
    \end{tabular}
    \end{adjustbox}
\end{table*}

\captionsetup[table]{aboveskip=0pt, belowskip=10pt}
\begin{table}[t]
    \vspace{-8pt}
    \caption{Comparison of different systems on the simulated test set. RTF is measured on a single NVIDIA 4090 GPU. ``Params." denotes the number of trainable parameters.}
    \label{tab:rtf_comparison}
    \centering
    \begin{adjustbox}{max width=\columnwidth}
    \begin{tabular}{l c c c c}
        \toprule[1pt]
        \multicolumn{1}{c}{System}            & Type & WER~(\%)~$\downarrow$ & RTF~$\downarrow$ & Params. (M)~$\downarrow$ \\
        \midrule
        Noisy                                 & —    & 23.9                  & —                & —                        \\
        \midrule 
        FRCRN~\cite{zhao2022frcrn}            & D    & \textbf{5.9}          & 0.024            & 13.1                     \\
        MP-SENet~\cite{lu2023mp}              & D    & 9.3                   & 0.019            & \textbf{2.1}             \\
        \midrule 
        SELM~\cite{wang2024selm}              & G    & 22.2                  & 0.042            & 219.0                    \\
        AnyEnhance~\cite{zhang2025anyenhance} & G    & 13.0                  & 1.423            & 324.0                    \\
        FlowSE~\cite{wang2025flowse}          & G    & 12.4                  & 0.121            & 337.1                    \\
        LLaSE-G1~\cite{kang2025llase}         & G    & 14.6                  & 0.057            & 1072.9                   \\
        \midrule
        MeanFlowSE                            & G    & 8.5                   & \textbf{0.013}   & 40.7                     \\
        MeanFlowSE$_\text{VAE input}$         & G    & 18.4                  & \textbf{0.013}   & 40.7                     \\
        MeanFlowSE$_\text{FM(40)}$            & G    & 8.8                   & 0.042            & 40.7                     \\
        MeanFlowSE$_\text{FM(100)}$           & G    & 7.7                   & 0.086            & 40.7                     \\
        \bottomrule[1pt]
    \end{tabular}
    \end{adjustbox}
\end{table}

\subsection{Datasets \& Evaluation Metrics}
\label{ssec:datasets_and_evaluation_metrics}

\textbf{Training data.} All models are trained on the Interspeech 2020 DNS Challenge dataset~\cite{reddy2020interspeech}, including clean speech, noise, and room impulse responses (RIRs). During training, each sample is constructed by first randomly selecting a clean speech segment, a noise segment, and an RIR. The clean speech is convolved with the RIR to simulate reverberation and then mixed with noise at a signal-to-noise ratio (SNR) randomly sampled between $-10$~dB and $20$~dB. The resulting mixture is cropped into a 4-second segment, and all audio signals are resampled to 16~kHz.

\textbf{Test sets.} For acoustic evaluation, we adopt the Interspeech 2020 DNS Challenge blind test set, which includes three subsets: \textit{With Reverb}, \textit{Without Reverb}, and \textit{Real Recording}. This allows us to assess performance under both controlled and real-world acoustic conditions. We also generate a simulated test set using the same procedure as in training.

\textbf{Evaluation Metrics.} We evaluate systems along three complementary axes: (1) Acoustic quality using DNSMOS~\cite{reddy2021dnsmos}; (2) Semantic Preservation using word error rate (WER) computed by OpenAI’s Whisper-Large ASR model~\cite{radford2023robust}\footnote{\url{https://huggingface.co/openai/whisper-large-v3}}; (3) Efficiency using RTF measured on a single NVIDIA RTX 4090 GPU.

These metrics collectively reflect listening quality, recognition robustness, and computational efficiency, providing a balanced assessment of SE models. For each case, we perform 5 independent sampling runs and report the best result.

\subsection{Implementation Details}
\label{ssec:implementation_details}

\textbf{Model configuration.} We employ WavLM-Large~\cite{chen2022wavlm}\footnote{\url{https://huggingface.co/microsoft/wavlm-large}}~as the SSL Encoder. Features from all 24 transformer layers are fused through a trainable weighted sum with softmax normalization to form the noisy condition. For clean speech, we use the WaveVAE from KALL-E~\cite{zhu2024autoregressive,lei2022glowwavegan2highqualityzeroshot}~as the VAE encoder–decoder, pre-trained on 16~kHz waveforms, producing 256-dimensional latent representations at 25~Hz. 

The DiT serves as the backbone of MeanFlowSE, consisting of $N=8$ transformer layers, each with 8 attention heads, hidden size 512, and feed-forward dimension 2048. MeanFlow is configured following~\cite{geng2025mean}~with a flow ratio of 0.25. The two time steps $(r,t)$ are sampled from a log-normal distribution $(\mu=-0.4,\sigma=1.0)$. The model is trained to predict the average velocity field via autograd-based JVP formulation, with adaptive weighting parameters $\gamma=0.5$ and $c=10^{-3}$. Time embeddings use sinusoidal frequency encoding followed by a linear layer, and positional encoding employs standard 1D sinusoidal embeddings.

\textbf{Training configuration.} All models are trained for 200 epochs using the AdamW optimizer, with an initial learning rate of $1\times10^{-3}$ exponentially decayed by 0.99 per epoch. Gradient clipping with a maximum norm of 1.0 is applied. Training is conducted on 8 NVIDIA RTX 4090 GPUs.

\textbf{Baseline Systems.} We compare MeanFlowSE with SOTA SE models, including discriminative models~\cite{zhao2022frcrn,lu2023mp}, flow-matching-based models~\cite{wang2025flowse}, language-model-based approaches~\cite{wang2024selm,kang2025llase}, and masked transformer model~\cite{zhang2025anyenhance}. To evaluate design choices, we include two MeanFlowSE variants: (1) replacing SSL-based latents with noisy VAE latents, and (2) setting the flow ratio to zero ($r=t$), which reduces the MeanFlow objective to standard flow-matching.

\subsection{Results}
\label{ssec:results}

Table~\ref{tab:dnsmos_comparison} reports the DNSMOS scores of all systems on the Interspeech 2020 DNS Challenge blind test set across three scenarios: \textit{With Reverb}, \textit{Without Reverb}, and \textit{Real Recording}. Generative models consistently achieve higher perceptual scores than discriminative models, highlighting their advantage in enhancing speech naturalness and listening quality. Among them, the proposed MeanFlowSE attains SOTA or comparable results across all DNSMOS metrics.

Table~\ref{tab:rtf_comparison} further compares WER, RTF, and model size. MeanFlowSE achieves the lowest WER among generative baselines, indicating superior preservation of linguistic content. It also delivers the lowest RTF at 0.013, surpassing even some discriminative models in latency. In terms of parameter efficiency, MeanFlowSE contains only 40.7~M trainable parameters, the smallest among the evaluated generative models. This compact design, together with strong DNSMOS performance and low-latency inference, demonstrates the high efficiency of MeanFlow for one-step generative modeling in SE and highlights MeanFlowSE’s potential for real-time speech communication applications.

\subsection{Ablation Study}
\label{ssec:ablation_study}

We conduct ablation studies to examine the effects of the two key design choices in MeanFlowSE.

\textbf{SSL conditioning v.s. VAE conditioning.} Replacing SSL-based latents with noisy VAE representations leads to noticeable degradation in both DNSMOS and WER. This suggests that VAE latents, being noisier and less structured, provide weaker guidance, whereas SSL embeddings capture richer contextual and phonetic information, improving perceptual quality and semantic preservation. These results highlight the importance of leveraging high-quality SSL features for robust speech enhancement.

\textbf{MeanFlow v.s. standard flow-matching.} Substituting MeanFlow with standard flow-matching using 40 or 100 inference steps shows that the 40-step variant performs comparably in DNSMOS but with higher RTF, while 100 steps slightly improve DNSMOS at the cost of even greater latency. This confirms that MeanFlow achieves an effective trade-off between enhancement quality and inference efficiency.

\section{Conclusions}
\label{sec:conclusions}

In this paper, we propose \textbf{MeanFlowSE}, a one-step generative speech enhancement model built on MeanFlow and a condition of SSL embeddings. Experimental results demonstrate that it achieves SOTA-level acoustic and semantic preservation while maintaining compact model size, low-latency inference, and strong robustness under diverse acoustic conditions, highlighting its potential for real-time applications. Future work will focus on further improving speech quality, adapting the model for low-latency streaming processing, and extending the framework to full-band scenarios.

\small
\bibliographystyle{IEEEbib}
\bibliography{strings,refs}

\begin{thebibliography}{10}

\bibitem{wang2023tfgridnetmakingtimefrequencydomain}
Zhong-Qiu Wang, Samuele Cornell, Shukjae Choi, Younglo Lee, Byeong-Yeol Kim, and Shinji Watanabe,
\newblock ``Tf-gridnet: Making time-frequency domain models great again for monaural speaker separation,'' 2023.

\bibitem{lu2023mp}
Ye-Xin Lu, Yang Ai, and Zhen-Hua Ling,
\newblock ``Mp-senet: A speech enhancement model with parallel denoising of magnitude and phase spectra,''
\newblock {\em arXiv preprint arXiv:2305.13686}, 2023.

\bibitem{zhao2022frcrn}
Shengkui Zhao, Bin Ma, Karn~N Watcharasupat, and Woon-Seng Gan,
\newblock ``Frcrn: Boosting feature representation using frequency recurrence for monaural speech enhancement,''
\newblock in {\em ICASSP 2022-2022 IEEE international conference on acoustics, speech and signal processing (ICASSP)}. IEEE, 2022, pp. 9281--9285.

\bibitem{CDiffSE}
Yen-Ju Lu, Zhong-Qiu Wang, Shinji Watanabe, Alexander Richard, Cheng Yu, and Yu~Tsao,
\newblock ``Conditional diffusion probabilistic model for speech enhancement,'' 2022.

\bibitem{StorM}
Jean-Marie Lemercier, Julius Richter, Simon Welker, and Timo Gerkmann,
\newblock ``Storm: A diffusion-based stochastic regeneration model for speech enhancement and dereverberation,''
\newblock {\em IEEE/ACM Transactions on Audio, Speech, and Language Processing}, vol. 31, pp. 2724–2737, 2023.

\bibitem{richter2023sgmse}
Julius Richter, Simon Welker, Jean-Marie Lemercier, Bunlong Lay, and Timo Gerkmann,
\newblock ``Speech enhancement and dereverberation with diffusion-based generative models,'' 2023.

\bibitem{wang2024selm}
Ziqian Wang, Xinfa Zhu, Zihan Zhang, YuanJun Lv, Ning Jiang, Guoqing Zhao, and Lei Xie,
\newblock ``Selm: Speech enhancement using discrete tokens and language models,''
\newblock in {\em ICASSP 2024-2024 IEEE International Conference on Acoustics, Speech and Signal Processing (ICASSP)}. IEEE, 2024, pp. 11561--11565.

\bibitem{yao2025gensegenerativespeechenhancement}
Jixun Yao, Hexin Liu, Chen Chen, Yuchen Hu, EngSiong Chng, and Lei Xie,
\newblock ``Gense: Generative speech enhancement via language models using hierarchical modeling,'' 2025.

\bibitem{mu2025fromcontinous}
Zhaoxi Mu, Rilin Chen, Andong Li, Meng Yu, Xinyu Yang, and Dong Yu,
\newblock ``From continuous to discrete: Cross-domain collaborative general speech enhancement via hierarchical language models,'' 2025.

\bibitem{li2024masksrmaskedlanguagemodel}
Xu~Li, Qirui Wang, and Xiaoyu Liu,
\newblock ``Masksr: Masked language model for full-band speech restoration,'' 2024.

\bibitem{zhang2025anyenhance}
Junan Zhang, Jing Yang, ..., and Zhizheng Wu,
\newblock ``Anyenhance: A unified generative model with prompt-guidance and self-critic for voice enhancement,''
\newblock {\em IEEE Transactions on Audio, Speech and Language Processing}, vol. 33, pp. 3085--3098, 2025.

\bibitem{zheng2025onestepdiffusion}
Bowen Zheng and Tianming Yang,
\newblock ``Revisiting diffusion models: From generative pre-training to one-step generation,'' 2025.

\bibitem{lipman2023flowmatchinggenerativemodeling}
Yaron Lipman, Ricky T.~Q. Chen, Heli Ben-Hamu, Maximilian Nickel, and Matt Le,
\newblock ``Flow matching for generative modeling,'' 2023.

\bibitem{wang2025flowse}
Ziqian Wang, Zikai Liu, Xinfa Zhu, Yike Zhu, Mingshuai Liu, Jun Chen, Longshuai Xiao, Chao Weng, and Lei Xie,
\newblock ``Flowse: Efficient and high-quality speech enhancement via flow matching,''
\newblock {\em arXiv preprint arXiv:2505.19476}, 2025.

\bibitem{yin2024onestepdiffusiondistributionmatching}
Tianwei Yin, Michaël Gharbi, Richard Zhang, Eli Shechtman, Fredo Durand, William~T. Freeman, and Taesung Park,
\newblock ``One-step diffusion with distribution matching distillation,'' 2024.

\bibitem{xu2025robustonestepspeechenhancement}
Liang Xu, Longfei~Felix Yan, and W.~Bastiaan Kleijn,
\newblock ``Robust one-step speech enhancement via consistency distillation,'' 2025.

\bibitem{geng2025mean}
Zhengyang Geng, Mingyang Deng, Xingjian Bai, J~Zico Kolter, and Kaiming He,
\newblock ``Mean flows for one-step generative modeling,''
\newblock {\em arXiv preprint arXiv:2505.13447}, 2025.

\bibitem{peebles2023scalable}
William Peebles and Saining Xie,
\newblock ``Scalable diffusion models with transformers,''
\newblock in {\em Proceedings of the IEEE/CVF international conference on computer vision}, 2023, pp. 4195--4205.

\bibitem{chen2019odesolver}
Ricky T.~Q. Chen, Yulia Rubanova, Jesse Bettencourt, and David Duvenaud,
\newblock ``Neural ordinary differential equations,'' 2019.

\bibitem{vectokspeech}
Xinfa Zhu, Yuanjun Lv, Yi~Lei, Tao Li, Wendi He, Hongbin Zhou, Heng Lu, and Lei Xie,
\newblock ``Vec-tok speech: Speech vectorization and tokenization for neural speech generation,''
\newblock {\em IEEE Transactions on Audio, Speech and Language Processing}, vol. 33, pp. 1243--1254, 2025.

\bibitem{kang2025llase}
Boyi Kang, Xinfa Zhu, Zihan Zhang, Zhen Ye, Mingshuai Liu, Ziqian Wang, Yike Zhu, Guobin Ma, Jun Chen, Longshuai Xiao, et~al.,
\newblock ``Llase-g1: Incentivizing generalization capability for llama-based speech enhancement,''
\newblock {\em arXiv preprint arXiv:2503.00493}, 2025.

\bibitem{reddy2020interspeech}
Chandan~KA Reddy, Vishak Gopal, Ross Cutler, Ebrahim Beyrami, Roger Cheng, Harishchandra Dubey, Sergiy Matusevych, Robert Aichner, Ashkan Aazami, Sebastian Braun, et~al.,
\newblock ``The interspeech 2020 deep noise suppression challenge: Datasets, subjective testing framework, and challenge results,''
\newblock {\em arXiv preprint arXiv:2005.13981}, 2020.

\bibitem{reddy2021dnsmos}
Chandan~KA Reddy, Vishak Gopal, and Ross Cutler,
\newblock ``Dnsmos: A non-intrusive perceptual objective speech quality metric to evaluate noise suppressors,''
\newblock in {\em ICASSP 2021-2021 IEEE International Conference on Acoustics, Speech and Signal Processing (ICASSP)}. IEEE, 2021, pp. 6493--6497.

\bibitem{radford2023robust}
Alec Radford, Jong~Wook Kim, Tao Xu, Greg Brockman, Christine McLeavey, and Ilya Sutskever,
\newblock ``Robust speech recognition via large-scale weak supervision,''
\newblock in {\em International conference on machine learning}. PMLR, 2023, pp. 28492--28518.

\bibitem{chen2022wavlm}
Sanyuan Chen, Chengyi Wang, Zhengyang Chen, Yu~Wu, Shujie Liu, Zhuo Chen, Jinyu Li, Naoyuki Kanda, Takuya Yoshioka, Xiong Xiao, et~al.,
\newblock ``Wavlm: Large-scale self-supervised pre-training for full stack speech processing,''
\newblock {\em IEEE Journal of Selected Topics in Signal Processing}, vol. 16, no. 6, pp. 1505--1518, 2022.

\bibitem{zhu2024autoregressive}
Xinfa Zhu, Wenjie Tian, and Lei Xie,
\newblock ``Autoregressive speech synthesis with next-distribution prediction,''
\newblock {\em arXiv preprint arXiv:2412.16846}, 2024.

\bibitem{lei2022glowwavegan2highqualityzeroshot}
Yi~Lei, Shan Yang, Jian Cong, Lei Xie, and Dan Su,
\newblock ``Glow-wavegan 2: High-quality zero-shot text-to-speech synthesis and any-to-any voice conversion,'' 2022.

\end{thebibliography}

\end{document}